# Structural and Magnetic Study of Metallo-Organic YIG Powder Using 2-ethylhexanoate Carboxylate Based Precursors

S. Hosseinzadeh[a], P. Elahi[b], M. Behboudnia[*, a], M.H. Sheikhi[c], S.M. Mohseni[*, d]

[a]Department of Physics, Urmia University of Technology, Urmia, Iran

[b]Department of Material Science, The University of Utah, Utah, U.S.

[c]Department of Communications and Electronics, School of Electrical and Computer Engineering, Shiraz, Iran

[d]Faculty of Physics, Shahid Beheshti University, Evin, 19839, Tehran, Iran

**Abstract**

The crystallization and magnetic behavior of yttrium iron garnet (YIG) prepared by metallo-organic decomposition (MOD) method are discussed. The chemistry and physics related to synthesis of iron and yttrium carboxylates based on 2-ethylhexanoic acid (2EHA) are studied, since no literature was found which elucidates synthesis of metallo-organic precursor of YIG in spite of the literatures of doped YIG samples such as Bi-YIG. Typically, the metal carboxylates used in preparation of ceramic oxide materials are 2-ethylhexanoate (2EH) solvents. Herein, the synthesis, thermal behavior and solubility of yttrium and iron 2EH used in synthesis of YIG powder by MOD are reported. The crystallization and magnetic parameters, including saturation magnetization and coercivity of these samples, smoothly change as a function of the annealing temperature. It is observed that high sintering temperature of 1300 to 1400 °C promotes the diffraction peaks of YIG, therefore, we can conclude that the formation of YIG in MOD method increases the crystallization temperature. The maximum value of saturation magnetization and minimum value of coercivity and remanence are observed for the sample sintered at 1200°C which are 13.7 emu/g, 10.38 Oe and 1.5 emu/g, respectively. This study cites the drawbacks in chemical synthesis of metallo-organic based YIG production.

*Corresponding author. E-mail Address: m-mohseni@sbu.ac.ir, majidmohseni@gmail.com (Seyed Majid Mohseni). mbehboudnia@gmail.com (Mahdi Behboudnia)

1. Introduction

Magnetic thin film of Yttrium Iron Garnet ($Y_3Fe_5O_{12}$ − YIG) has found great attention in emerging spintronic devices. Such a thin film with low Gilbert damping constant is suitable for magnonics and beside its insulator behavior (insulating nature), it gains a great deal of attention in generation of spin current. With the advent of spintronics, the demand for synthesis and processing of YIG films has surged forward greatly.

In general, the characteristics and surface morphology of the thin films strongly depends on deposition techniques. Pulsed laser deposition (PLD) has emerged as a preferred technique to deposit complex oxide thin films, heterostructures, and superlattices with high quality in comparison with the other deposition techniques such as e-beam evaporation and sputtering [4, 5]. Contrarily, chemical solution deposition (CSD) techniques are cost effective synthesis processes in which the precursor solution is deposited by variety of relatively simple techniques such as spin or dip coating followed by post treatments of drying and annealing in case needed. More recently, metallo-organic decomposition (MOD) recognized as a chemical technique has been growing due to its extremely good molecular level composition controllability in the fabrication process of garnet thin films. Kirihara et.al [6] reported generation of spin-current-driven thermoelectric conversion by using Bi-YIG thin layer prepared by MOD. There are also further literatures [7-9] which report deposition of doped YIG and YIP (yttrium iron perovskite, $YFeO_3$) by MOD technique using metal carboxylate precursors in organic solvents. In majority of the publications, the effect of the annealing temperature and duration of doped YIG powders has been investigated on the physical and magnetic properties. It is shown that at higher temperatures, the observed physical and magnetic properties of garnet are similar to that of solid state reaction techniques [10].

The MOD procedure introduced by Azevedo et. al is used for thin film deposition [11-14]. Their procedure consists of preparation of metal-organic carboxylates, which exhibit high compatibility to organic solvents, result in better final thin film properties. They succeeded in preparing polycrystalline $Gd_2BiFe_5O_{12}$ and $(DyBi)_3(Fe, Ga)_5O_{12}$ thin films on glass substrates. In their work, large carboxylate compounds of 2-ethylhexanoate (2EH) precursors were synthesized, and then a stoichiometric ratio of precursors was dissolved in xylene followed by spin coating of the solution on preferred substrate.

Typically, the metal carboxylates used in the preparation of ceramic oxide materials are 2EH salts which are dispersed in aromatic solvents [10, 12, 13, 15, 16]. Metal 2EH have found wide application such as metal–organic precursors, catalysts for ring opening polymerizations and drier agent in paint industries [17]. Realization of the MOD advantages requires in-depth understanding of the precursor solution chemistry such as precursor species, solute concentration, and solvent system and its relation to the final material properties. A detailed study of the fundamental properties of 2EH yttrium 2-ethylhexanoate (Y-2EH) and iron-2ethylhexanoate (Fe-2EH) precursors used in synthesis of YIG thin films has not been reported yet. The primary objective of this work is to investigate the fundamental properties of yttrium 2-ethylhexanoate (Y-2EH) and iron-2-ethylhexanoate (Fe-2EH) precursors needed to synthesis

YIG thin films, such as chemical properties (including solvent/solute concentration ratios, solution structure and internal bonding) and on process optimization methodologies in order to obtain optimum YIG properties.

## 2. Materials and methods

### 2.1. Reagents

Iron(III) nitrate nonahydrate ($Fe(NO_3)_3$), Yttrium nitrate hexahydrate ($Y(NO_3)_3$), $Y_2O_3$, 2EHA, Xylene, Toluene, n-Hexane, n-octane, benzene e, THF, Isopropanol, propionic acid, and glacial acid acetic are of analytical reagent grade.

### 2.2. Synthesis of Yttrium and Iron carboxylates

A metal carboxylate compound is defined as the central metal atoms linked to ligands through a hetero –atoms [14].

#### 2.2.1 Fe-2EH:

The iron carboxylates were prepared by double decomposition from ammonium soaps obeying the following reactions [18]

$NH_4OH + C_7H_{15}COOH_1$ 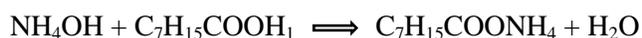 $C_7H_{15}COONH_4 + H_2O$

$Fe(NO_3)_3 + 3\ C_7H_{15}COONH_4$ 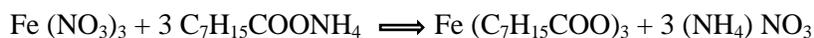 $Fe(C_7H_{15}COO)_3 + 3\ (NH_4)NO_3$

The first reaction involves the preparation of the ammonium soap of 2-ethylhexanoic acid (2EHA). Then the soap from second reaction was mixed with the aqueous solution of $Fe(NO_3)_3$. After stirring for 10 minutes, the solution was separated into $Fe(RCOO)_3$ and $(NH_4)NO_3$. In a funnel separator, a solvent (xylene) was added to this solution and the carboxylate which dissolves in xylene was separated and filtered by 0.22 µ-filter and was kept away until xylene evaporated under fume hood to reach a reddish brown powder of Fe-2EH. By decomposing a certain amount of the Fe-2EH into the metal oxide and weighing the oxide, the equivalent amount of iron oxide in Fe-2EH was determined.

**2.2.2 Y-2EH**: The Y-2EH is prepared according to the following reaction [19]

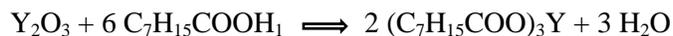

$Y_2O_3 + 6\ C_7H_{15}COOH_1 \implies 2\ (C_7H_{15}COO)_3Y + 3\ H_2O$

Yttrium oxide ($Y_2O_3$) powder was added gradually to 2EHA while gently stirring and it was kept at 100°C till all liquid has been evaporated. The product was stirred with an excess of toluene for 24h at room temperature. Thereupon, it was filtered and was kept under fume hood until the toluene was evaporated and a white solid is achieved. This white solid is Y-2EH. By decomposing a known amount of the Y-2EH into the metal oxide and weighing the oxide, the equivalent amount of $Y_2O_3$ in Y-2EH was determined.

### 2.3. Synthesis of YIG powder

Metallo-organic YIG was prepared using Y-2EH and Fe- 2EH in stoichiometric ratio of 3:5. Firstly, The Y- 2EH and Fe-2EH were dissolved into toluene and xylene, respectively. Secondly, the solutions were mixed together with respect to the weight percentage of each carboxylate to achieve the desired stoichiometry. The process was followed by adding glacial acetic acid until a homogeneous solution was reached with no precipitation. The powders of MOD solution were prepared by drying for 72 hours at 150 °C and then they were grinded in a mortar. Figure 1 illustrates the schematic diagram of metallo-organic decomposition of YIG.

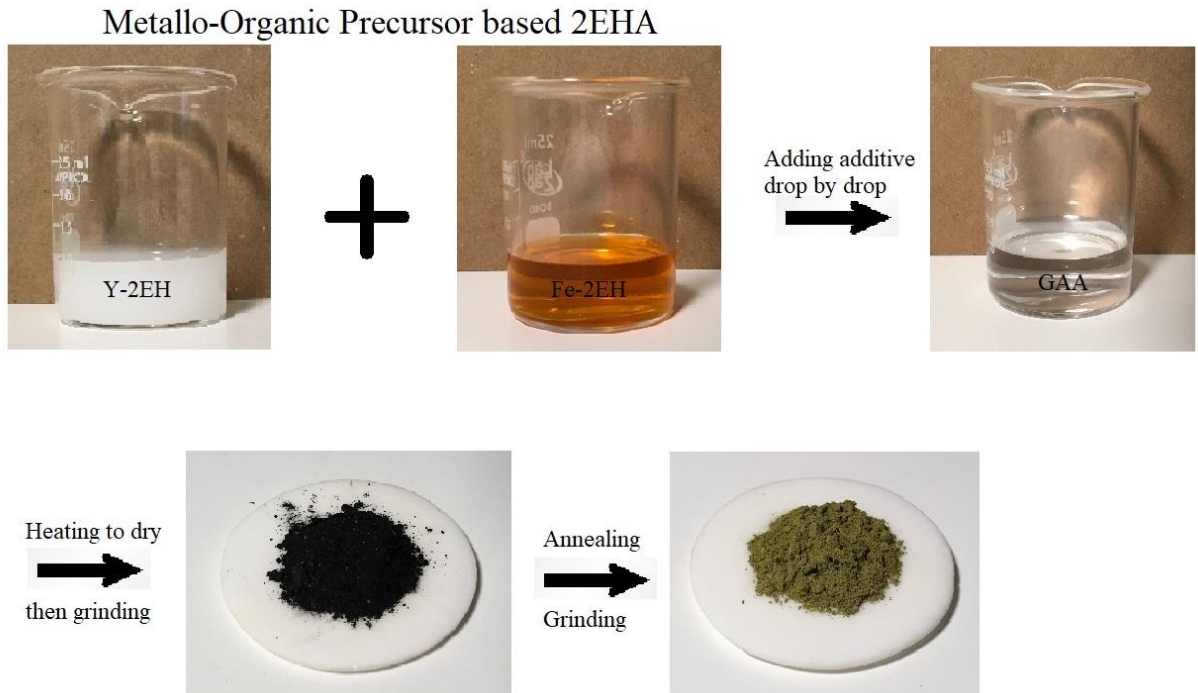

**Figure 1:** Schematic diagram for the metallo-organic decomposition process of YIG.

### 2.4. Characterization of samples

To investigate the pyrolysis and crystallization process of the YIG prepared by the MOD method, a thermo-gravimetric-differential thermal analysis (TGA-DTG-DTA) (model mettle Toledo C1600 analyzer) was carried out from controlled room temperature to 1400°C in an air atmosphere with the heating rate of 10°C/min. The crystalline structure of samples was characterized using X-ray diffractometer (STOE-STADI) with Cu Kα ($\lambda$ = 0.154 nm) radiation. Room temperature magnetization measurements were performed using vibrating sample magnetometer (Meghnatis Daghigh Kavir Co.)

### 3. Results and discussions

As previously reported by Beckel [20] and Neagu [21], the solvent influences the boiling point of the solution and determines the speed of evaporation during heating of the droplets which affects the film roughness. The solvent also impacts the maximum metal carboxylate solubility and spreading behavior of the droplets. The deposition temperature is primarily influenced by the solvent and additives. These organic solvents and additives help in gelation and polymerizations, and modification of the solution properties [21-24], such as viscosity, solubility of metal carboxylate and spreading of the droplets. The propionic acid and mixture of ethanol and 2EHA were used as additives as mentioned in literatures beyond the number. Besides, glacial acetic acid (GAA) were used as additive. This study reveals that GAA improves YIG formation and decreases YIG crystallization temperature.

Figure 2 demonstrates the TGA-DTG and DTA curves of the MOD precursor. It shows thermal decomposition proceeds via six-step processes in an air atmosphere. The broad endothermic peak from room temperature to about 200 °C with a total weight loss of about 3% corresponds to the evaporation of residual solvents including xylene with boiling point ($b_p$) of 138 °C and glacial acetic acid with boiling point of 118 °C [25]. Three exothermic peaks from 200 to about 480 °C with total weight loss of 51.5 % are ascribed to the volatilization of 2EHA (bp~228 °C) and the pyrolysis of possible metal-organic compounds such as Fe-2EH are expected [25, 26]. The three exothermic peaks from 480 to 820 °C represent removal of the three 2-EH groups from the Y-2EH molecule to form $Y_2O_3$ with total weight loss of 20.7% [27]. The following two exothermic peaks at temperature range of 820 to 920 °C with a weight loss of 2.6% correspond to crystallization of YIG and YIP. The weight loss curve then approaches plateau from temperature range 920 to 1300°C. The two exothermic peaks observed within this temperature range are attributed to the conversion of YIP to YIG and crystallization of YIG. This crystallization temperature is higher than other chemical solution decomposition methods [28, 29]. On the other hand, the main advantage of the carboxylate-based-routines is the comparably low crystallization temperature. This is due to the educt molecules that are mixed at the molecular level. Thus, the diffusion paths of metal-and-oxygen-ions are short compared to classical powder-based syntheses of ceramic bulk materials [18, 30]. However, The formation of YIG in the MOD method increases the crystallization temperature which is mentioned previously by Lee et al. [13]. The above results suggests that the 2EHA may not serve as an excellent ligand for yttrium precursor, since the decomposition of the organics for Y-2EH occurs at temperatures higher than 500°C as reported in literatures [27].

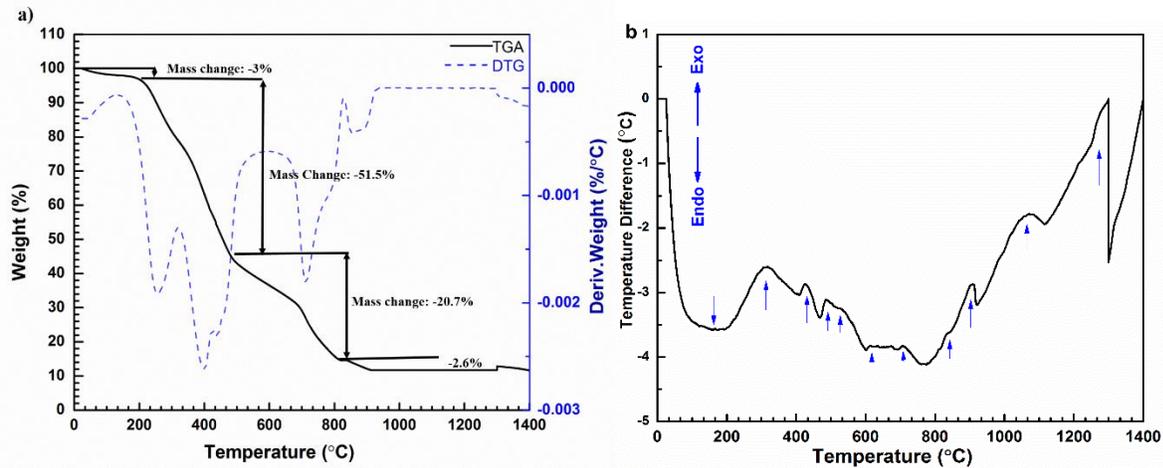

**Figure 2**: TGA-DTG (a), DTA (b) characteristics for YIG powder

Figure 3 shows the XRD patterns of YIG particles annealed at 1000-1400 °C for 2hrs. The XRD pattern of the YIG particles calcined at 1000 °C is associated with the formation of YIG with already formed yttrium oxide ($Y_2O_3$) around $2\theta=29°$ and hematite ($\alpha$-$Fe_2O_3$) around $2\theta=33°$ as the major phases in addition to the traces of YIP around $2\theta=48, 33°$. As the solid state reaction method suggested, the crystallization process could be described by the following equations,

$Y_2O_3 + Fe_2O_3 \Longrightarrow 2YFeO_3$ at 800-1200 °C

$3YFeO_3 + Fe_2O_3 \Longrightarrow Y_3Fe_5O_{12}$ at 1000-1300 °C

which indicates that the $Y_2O_3$ phase would first transform into YIP phase at low temperatures and then converts to YIG phase by combining with α-$Fe_2O_3$ at higher temperatures. The color of the particles calcined at 1000 °C temperature was reddish brown which is due to the presence of α-Fe2O3 and YIP as the major phases. When the YIG particles were calcined at 1100 °C and 1200°C, the color converted to brownish green, indicating the conversion of YIP to YIG phase. For the sample calcined at these two temperatures, the diffraction peak around 2θ=33° became stronger, the $Y_2O_3$ and YIP peaks became weaker, but still there is excess amount of α-$Fe_2O_3$ phase. As the annealing temperature is increased to 1300 & 1400 °C there is no change in diffraction peaks of YIG and α-$Fe_2O_3$ and the peak related to the α-$Fe_2O_3$, located near 2θ= 33°, still remains the same and the intensity of garnet peaks are not increased as sintering temperature is risen. The remained α-$Fe_2O_3$ phase can be explained due to the insufficient amount of Y-2EH during the reaction because of precipitation of Y-precursor. The metal carboxylate must be soluble and stable at room temperature in the solution, if not precipitation will occur and lead to inhomogeneity repartition of cations in the obtained gel.

The successful application of the MOD process significantly depends on the metallo-organic compounds used as precursors for a variety of elements. The ideal compounds should satisfy some requirements such as high solubility in a common solvent. The solutions of individual metallo-organic compounds should mix in the appropriate ratio to give the desired stoichiometry for final formation. The main conclusion is that there is no theoretical database for selecting the optimum solvent suitable for MOD process. In order to explore the interactions between solute and solvent system, the polarity of the solvent and solute should be considered due to evaluation of the effect of dipole-dipole interactions.

Generally, to have a good solubility, the polarity of solute and solvent should be close to each other. The longer chain acid like 2EHA can be solved in low polar solvents (eg. Xylene, alcohol etc.). In case of unknown solubility parameter of a compound, a successful approach is to first try a non-polar solvent which has low solubility parameter. If the approach is not successful, then a moderately polar solvent with intermediate solubility parameter should be tried.

In order to find an adequate solubility in the desired solvents which were compatible with each other, we tested some solvents recommended by literatures for yttrium and iron 2EH. Therefore, we tested xylene, toluene, benzene, n-hexane, THF for both Fe-2EH and Y-2EH and found that the homogeneity of toluene and xylene are the best for yttrium and iron, respectively. However, the solubility and homogeneity of Y-2EH tends to be much less than that of the Fe-2EH. The Y-2EH showed precipitation and was not as homogenous as Fe-2EH. As reported by Ishibashi et al. [12], the Y-2EH cannot dissolve in the solvent introduced by Azevedo et al.[14]. As a result, we suggest that synthesis of Y-2EH is not an easy and homogenous synthesis approach.

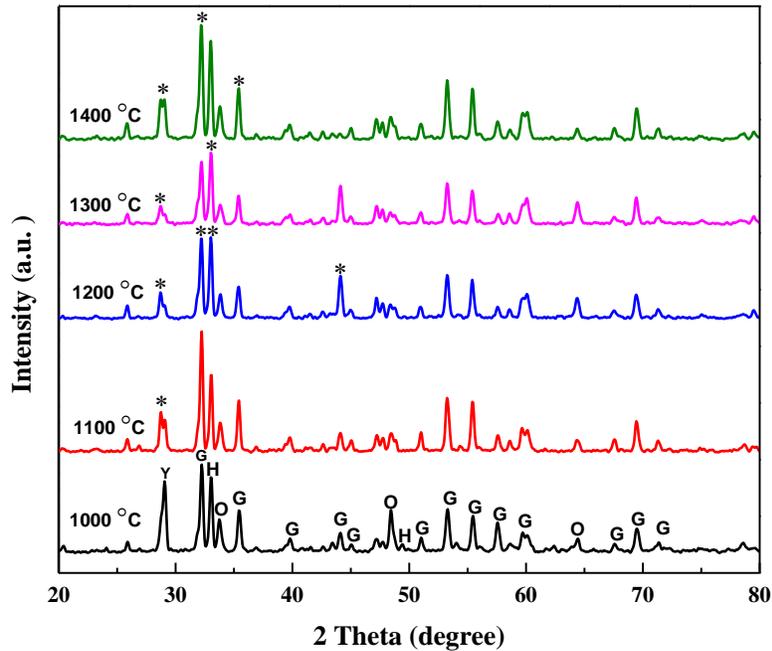

**Figure 3.** XRD pattern of the YIG powder annealed at 1000-1400 °C. Assignment of diffraction peaks are indicated as following: G: YIG, O: YIP, H: α-Fe2O3, Y: Y2O3

Figure 4 (a) shows the sintering temperature dependence of magnetization. Parameters such as $M_S$, $H_c$ and $M_r$ are shown in Figure 4 (b,c). The $M_S$ of the powders sintered at 1000-1400 °C were 9 to 13 emu/g, and a maximum value of 13.7 emu/g was observed for the powder sintered at 1200°C. From the XRD results, we observed that the intensity of garnet phases around 2θ=32, 45° are strongest at 1000-1400°C, which can be deduced that the magnetic behavior of sintered powders was strongly determined by the garnet phases due to the weak ferromagnetic properties of the $\alpha$-$Fe_2O_3$ and YIP phases. The magnetic results is similar to the magnetic results reported by Lee et al. [10]. In the range of 1000-1400°C, the $H_c$ and $M_r$ decrease whereas the $M_S$ shows a rise. The decrease in $M_r$ and $H_c$ versus the increase in $M_S$ are explained due to an increase in the particle size [31].

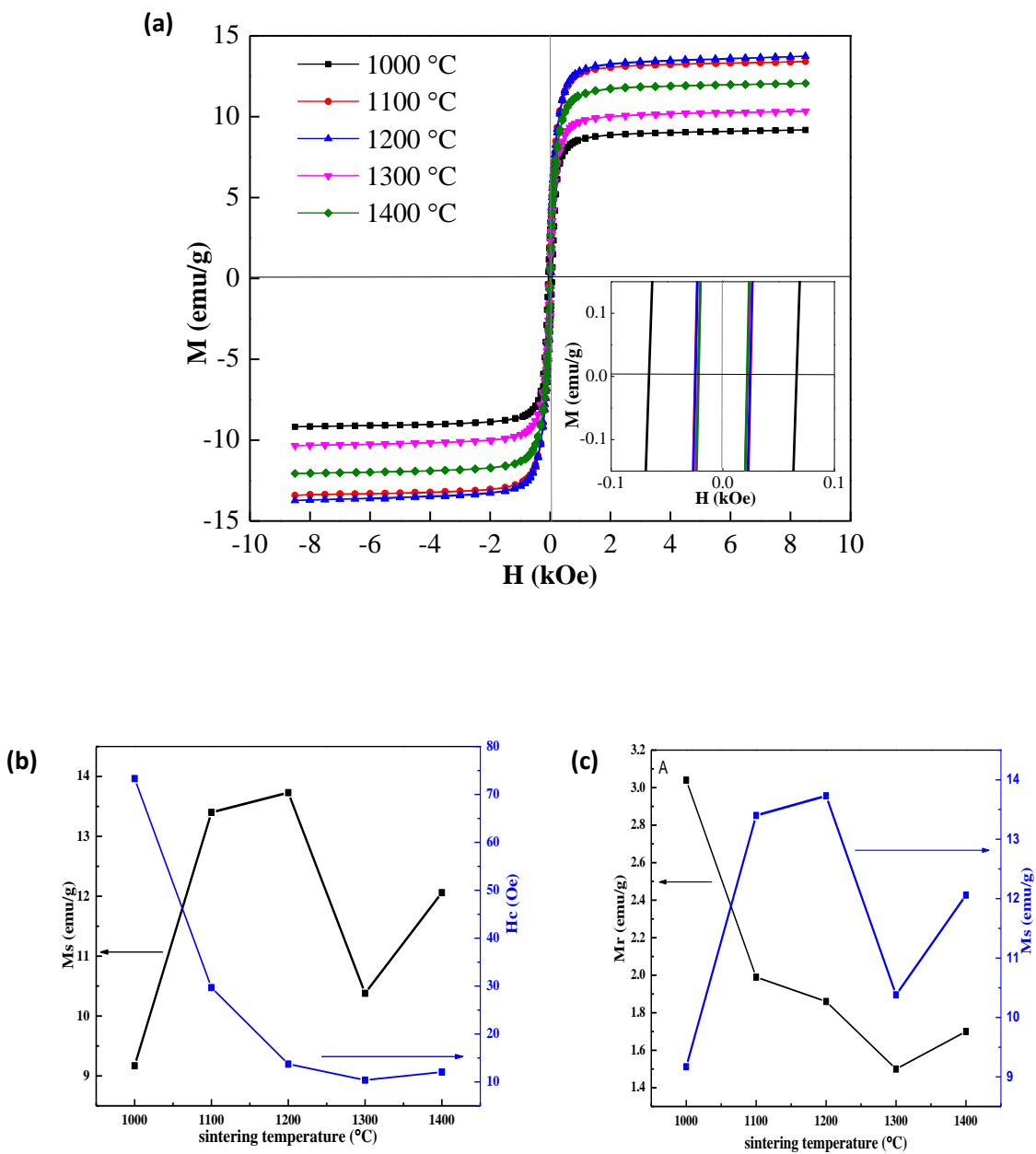

**Figure 4**. (a) Room temperature magnetization hysteresis loops of powders sintered at 1000-1400 °C, (b)Variation of $M_S$, $H_c$, and (c) $M_r$ of YIG powders as a function of sintering temperature

## 4-Conclusion:

Metallo-organic precursors of yttrium and iron metal-carboxylates were synthesized and the chemistry and physics related to various fabrication steps were investigated. The metallo-organic compound in work can be dissolved in proper solvents such as toluene and xylene with the GAA used as an additive, to achieve the desired stoichiometry for preparing the YIG powder. Crystallization and magnetic behavior of the YIG was studied. It is observed that the Y-2EH shows precipitation and is not as stable as Fe-2EH and also Y-2EH is not homogenously synthesized. Our results can be valuable to revive useful materials for chemical solution processing of YIG family thin films.

## Acknowledgments

S.M. Mohseni acknowledges support from Iran Science Elites Federation (ISEF), Iran Nanotechnology Initiative Council (INIC) and Iran's National Elites Foundation (INEF)